# Theoretical Economics and the Second-Order Economic Theory. What is it?

Victor Olkhov

Independent, Moscow, Russia

victor.olkhov@gmail.com

ORCID: 0000-0003-0944-5113

March 15, 2024

## ABSTRACT

The economic and financial variables of economic agents determine macroeconomic variables. Current models consider agents' variables that are determined by the sums of values and volumes of agents' trades during some time interval $\Delta$. We call them first-order economic variables. We describe how the volatilities and correlations of market trade values and volumes determine price volatility. We argue that such a link requests consideration of agents' economic variables of the second order that are composed of sums of squares of agents' transactions during $\Delta$. Almost any variable of the first order should be complemented by its second-order pair. Respectively, the sums of agents' second-order variables introduce macroeconomic variables of the second order. The description of the first- and second-order macroeconomic variables establishes the subject of second-order economic theory. We highlight that the complexity of second-order economic theory essentially restricts any hopes for precise predictions of price probability and, at best, could provide estimates of price volatility. That limits the predictions of price probability to Gauss's approximations only.

---

This research received no support, specific grants, or financial assistance from funding agencies in the public, commercial, or nonprofit sectors. We welcome valuable offers of grants, support, and positions.

## 1. Introduction

Studies in economic theory cover a wide range of research as general equilibrium and market design, asset pricing and behavioral economics, macroeconomics and monetary economics, decision and game theory, networks, etc. Studies by Richard Cantillon (1755), "An Essay on Economic Theory," by A. Cournot (1838), "Researches into the Mathematical Principles of the Theory of Wealth," and "Essentials Of Economic Theory" by J.B. Clark (1915) prove that economic theory is a subject of endless research. These were complemented by an enormous number of publications on different economic models - General Equilibrium (Neumann, 1945; Vines and Wills, 2018), DSGE (Christiano, Eichenbaum, and Trabandt, 2018), Economic Growth (Aghion and Durlauf, 2005), Game Theory (Samuelson, 2016), Behavioral Economics (Thaler, 2016), Maximum Principles in Economics (Samuelson, 1970), etc. Numerous models describe important relations that govern economic growth (Aghion and Durlauf, 2005), asset pricing and economic fluctuations (Cochrane, 2017), consumption (Friedman, 1957), and employment (Keynes, 1936), etc.

However, theoretical economics still has a huge stockpile of unsolved problems whose description is necessary for the management of real economic processes. A great number of economic variables reveal the complexity and diversity of economic and financial processes. Hence, it is important to study the general properties of the macroeconomic variables that describe economic evolution and consider them on an equal basis.

To do that, we simplify the problem and choose a few issues, the primary bricks of theoretical economics that can serve as a ground for the successive approximations of real economic processes. As such, we consider economic agents, agents' economic and financial variables, market trades that agents perform with other agents, and numerous agents' expectations that impact trade decisions and transactions with other agents. All these issues are well known and have been investigated for decades. We do not choose or highlight the description of any particular economic variable like consumption, demand, or investment. We consider the general properties of economic variables and market trades and study how agents' economic variables and market deals help describe macroeconomics and macro-finance.

Econometrics consider macroeconomic and macro-financial variables as composed of sums of corresponding variables of economic agents (Fox et al., 2017). In turn, the changes in agents' variables are determined by agents' trade values and volumes during a certain time interval *Δ*. Consumption and investment, supply and demand, etc., of agents can be determined by sums of values or volumes of trades between agents during a particular interval *Δ*. We call the theories that consider collective agents' variables that are composed by the sums of agents' trade values and volumes first-order economic theories. We refer to Fox et al. (2017) as a perfect presentation of methodology that describes methodologies based on the collection of agents' variables to obtain macroeconomic ones. However, these first-order economic theories are insufficient for an adequate description of economics.

The confirmation of that is simple. Economic agents make transactions in accordance with their expectations. The forecasts of market prices and predictions of price volatility play a core role in agents' expectations and trade decisions. Price volatility is the first obstacle that makes first-order economic theories insufficient. Below, we show that the forecasts of price volatility that are determined by the *2-d* price statistical moment require the predictions of the sums of squares of values and volumes of trades during the averaging interval *Δ*. We call the models that take into account the evolution of sums of squares of trade values and volumes second-order economic theories.

Respectively, the predictions of the *3-d* statistical moment of price should describe the sums of the third degrees of trade values and volumes and require the development of third-order economic theories. The imaginable exact predictions of price probability at horizon *T* need the development of economic theories that describe sums of *n-th* power of trade values and volumes during interval *Δ* for all *n=1,2,…*. That is completely impossible. Understanding the constraints between the predictions of price probability and the description of sums of the *n-th* degree of trade values and volumes may help develop successive approximations. However, no miracles or simple solutions to that problem exist.

We skip almost all technical details and math equations, which form the essence of our theoretical models, and explain the necessity, imminency, and meaning of the second-order economic theory in a most simple manner.

## 2. Theoretical economics

Theoretical economics uses numerous different notions, variables, and relations. We refer to the Essays by Richard Cantillon (1755) as absolutely contemporary research in economic theory. Due to the efforts of M. Thornton and C. Saucier, one can enjoy Cantillon's study, and even its content highlights all the main economic notions such as international trade and business cycles, market prices, money, interest, etc. References above prove that these important economic variables and processes have been under investigation until now.

Let us raise some questions. Are there any general relations and laws that govern the evolution of economic variables? How can one describe the factors that define the state and evolution of economic and financial variables? What are the successive approximations that can describe economic evolution? Studies of these problems can help model economic relations between particular variables like supply and demand, investment, economic growth, etc. Let us consider these questions most simply.

1. First, let us define simple elements of the economic system. We consider all participants in economic relations as economic agents. Economic agent-based models have been known for decades (Tesfatsion and Judd, 2006; Hamill and Gilbert, 2015). We regard economic agents as primary units, the simple bricks of economics. Different agents can perform different economic roles and have different economic variables. We consider corporations and banks, industry plants and households, travel agencies, government entities, etc., as economic agents. Economic and financial variables of these agents – credits and investment, assets under management, turnover, profits, and consumption, labor and taxes, etc. - define the origin of all macroeconomic and macro-financial variables. Aggregations of the economic and financial variables of all agents of the economy define macroeconomic and macro-financial variables that describe the economic state and evolution (Fox et al., 2017). The description of agents' economic and financial variables helps model and forecast macroeconomic variables.

2. Agents' variables change due to market transactions and deals performed with other agents. Investment and credit deals, market trades with assets or currencies, and transactions of commodities and energy between agents define the change in economic activity. Leontief (1955; 1973) collected transactions between productive sectors and industries as a tool for economic modeling. Trades between agents are the only factors that change agents' variables. Consumption and investment, purchasing and taxes, international trade, and labor appear as various forms of transactions and trades between agents. The descriptions of the deals between agents model the evolution of agents' variables and, therefore, model macroeconomic evolution as well.

3. Agents make economic and financial transactions and trades. In turn, trade decisions are based on agents personal expectations. Agents develop their expectations based on the economic and market environment, the expectations of other agents, and other "to-day" and "next-day" factors that impact market prices, economics and finance, inflation and currency rates, etc. Agents' expectations (Muth, 1961; Manski, 2017; Farmer, 2019) as the drivers of agents' trade decisions and the drivers of market activity have been studied for decades, and the above references only indicate the importance of expectations for adequate economic modeling.

We consider macroeconomic and macro-financial variables that design the economic state and evolution; agents' economic and financial variables that compose macro-variables; trades between agents that are the only factors that change agents' variables; and numerous agents' expectations that force agents to make deals as the basis and "elementary bricks" of theoretical economics. We consider the descriptions of agents' variables, transactions, and expectations as the key problems of theoretical economics.

## 3. First-order economic theory

One can consider present economic models as first-order economic theories that describe relations between macroeconomic and macro-financial variables, which are composed as sums of agents' economic variables. These agents' variables, in turn, are composed of sums of the corresponding market trades between agents during some time interval $\Delta$. For example, the investment made by an agent during the time interval $\Delta$ is determined as the sum of the investment deals made by this agent during that particular time term $\Delta$. The agent's consumption during $\Delta$ is determined by the sum of all consumption deals of this agent during $\Delta$. Such a time interval $\Delta$ can be equal to a month, quarter, year, etc. The duration of $\Delta$ plays a key role in theoretical economics, and different $\Delta$ can result in different descriptions of economic processes. The common property of the first-order models is that agents' variables are composed of sums of market trades and transactions between agents during $\Delta$.

However, economics and finance also depend on economic variables that can't be presented as sums of trades between agents. Price volatility is an example of one of the most influential factors that impacts markets, macro-finance, and macroeconomic evolution. The problem of price volatility predictions creates the next level of complexity in theoretical economics. We underline the impact of such a "particular" variable as price volatility spreads from the modeling of the financial market onto the description of the entire macroeconomics. In the next section, we consider the complexities generated by the modeling of market-based price volatility.

## 4. Market-based price volatility

Let us consider the market trades with some assets. Each market deal at a moment $t_i$ is described by the trade value $C(t_i)$, volume $U(t_i)$ and price $p(t_i)$ determined by a primitive equation:

$$C(t_i) = p(t_i)U(t_i) \quad (4.1)$$

Time series of market deals can define the trade values $C(t_i)$, volumes $U(t_i)$, and prices $p(t_i)$ with high frequencies, and usually, market price $p(t_i)$ at moment $t_i$ is described as a mean price $p(t)$ at time $t$ determined during some averaging time interval $\Delta$ that can be equal to an hour, day, week, etc. The definition of mean price $p(t)$ depends on the choice of price probability during interval $\Delta$. This "simple" issue hides a lot of complexities (Olkhov, 2021; 2022). Indeed, usually it is assumed that price $p(t_i)$ time series at the moments $t_i$ during $\Delta$ (4.2):

$$\Delta = \left[t - \frac{\Delta}{2}, t + \frac{\Delta}{2}\right] \quad ; \quad t_i \in \Delta\,, \; i = 1, \dots N \quad (4.2)$$

define frequency-based price probability $f(p_k)$ and price *n-th* statistical moments as follows:

$$f(p_k) \sim \frac{1}{N} m(p_k) \quad ; \quad E[p^n] \sim \frac{1}{N} \sum_k p_k^n \, m(p_k) = \frac{1}{N} \sum_{i=1}^{N} p^n(t_i) \quad (4.3)$$

Here $m(p_k)$ defines the number of deals at a price $p_k$. The definition of the price probability $f(p_k)$ (4.3) for $N$ trades performed during $\Delta$ is the common way to assess the market price probability (Shiryaev, 1999). The symbol ~ highlights that the finite number $N$ of trades gives only an assessment of probability. However, time series of the trade values $C(t_i)$ and volumes $U(t_i)$ during $\Delta$ (4.2) also define frequency-based probabilities of trade value $\nu(C_k)$ and volume $\mu(U_k)$:

$$\nu(C_k) \sim \frac{1}{N} \, m(C_k) \quad ; \quad \mu(U_k) \sim \frac{1}{N} \, m(U_k) \quad (4.4)$$

Here $m(C_k)$ defines the number of deals at value $C_k$ and $m(U_k)$ is the number of deals at volume $U_k$. Actually, the equation (4.1) prohibits the independent definitions of probabilities of the trade value $C(t_i)$, volume $U(t_i)$, and price $p(t_i)$. Given the probabilities $\nu(C)$ and $\mu(U)$ of the trade value $C(t_i)$ and volume $U(t_i)$ (4.4), the equation (4.1) should define the probability $f(p)$ of the market price $p(t_i)$, and it could be different from (4.3).

It is well known that a random variable can be described equally by a probability measure, characteristic function, and set of the *n-th* statistical moments (Shephard, 1991; Shiryaev, 1999). The number $M$ of the *n-th* statistical moments, $n=1,..M$, determines the *M-th* approximation of the probability measure. The more statistical moments are described, the more precise an approximation of the probability can be obtained. The first two statistical moments present a simple Gaussian approximation of a probability that depends on the average and volatility of the random variable.

Due to a finite number $N$ of market deals during $\Delta$ (4.2), one can assess only a finite number of market trade statistical moments $C(t;n)$ of the trade values and $U(t;n)$ (4.5) of the trade volumes that approximate the corresponding probabilities $\nu(C)$ and $\mu(U)$:

$$C(t;n) = E[C^n(t_i)] \sim \frac{1}{N} \sum_{i=1}^{N} C^n(t_i) \;\; ; \; U(t;n) = E[U^n(t_i)] \sim \frac{1}{N} \sum_{i=1}^{N} U^n(t_i) \qquad (4.5)$$

We use statistical moments *C(t;n)* and *U(t;n)* (4.5) of market trade to derive market-based statistical moments of price and refer to Olkhov (2021; 2022) for further details. The market-based average price *a(t;1)* (4.6) has been well-known for 40 years since a paper by Berkowitz et al. (1988) as volume weighted average price (VWAP). We denote the market-based average price *a(t;1)* (4.6) to highlight the distinctions from the statistical moments that are estimated by the frequency-based price probability (4.3). However, almost the same notion was implicitly introduced by Markowitz (1952), more than 35 years before the paper by Berkowitz et al. (1988). Indeed, Markowitz defines the return of the portfolio as the return weighted average by the "relative amount $X_i$ invested in security *i*" (Markowitz, 1952). Actually, each investor makes almost the same calculation to assess the average price of assets in his portfolio as the ratio of their total value to their total volume. One can consider market trades during *Δ* (4.2) as a "portfolio" and obtain exactly the definition of VWAP:

$$a(t;1) = \frac{\sum_{i=1}^{N} p(t_i)U(t_i)}{\sum_{i=1}^{N} U(t_i)} = \frac{C_\Sigma(t;1)}{U_\Sigma(t;1)} = \frac{C(t;1)}{U(t;1)} \qquad (4.6)$$

The relations (4.5; 4.7) determine the total value $C_\Sigma(t;n)$ and volume $U_\Sigma(t;n)$ of the *n-th* degrees of deals during *Δ* (4.2):

$$C_\Sigma(t;n) = \sum_{i=1}^{N} C^n(t_i) = N \cdot C(t;n) \quad ; \quad U_\Sigma(t;n) = \sum_{i=1}^{N} U^n(t_i) = N \cdot U(t;n) \quad (4.7)$$

The portfolio's return by Markowitz (1952) and VWAP (4.6) both give estimates of the average return and average price in a manner different from the frequency-based assessments *E[p]* (4.3). We denote VWAP as market-based average price. To obtain market-based price volatility, one should consider the squares of the trade price equation (4.1):

$$C^2(t_i) = p^2(t_i)U^2(t_i) \qquad (4.8)$$

In Olkhov (2021; 2022), we show that (4.8) generates the average *p(t;2)* (4.9) of the squares of price in the form that is similar to the definition of VWAP (4.5-4.7):

$$p(t;2) = E[p^2(t_i)] = \frac{\sum_{i=1}^{N} p^2(t_i)U^2(t_i)}{\sum_{i=1}^{N} U^2(t_i)} = \frac{C_\Sigma(t;2)}{U_\Sigma(t;2)} = \frac{C(t;2)}{U(t;2)} \qquad (4.9)$$

However, (4.9) can't serve as the 2-d market-based price statistical moment. To define the 2-d market-based price statistical moment *a(t;2)* (4.13), one should require the coherence of the market-based statistical moments *a(t;1)* and *a(t;2)* that result in the necessary inequality (4.10) for the market-based price volatility $\sigma^2(t)$:

$$\sigma^2(t) = a(t;2) - a^2(t;1) \geq 0 \qquad (4.10)$$

The relations (4.6; 4.9; 4.10) permit introduce the market-based price volatility $\sigma^2(t)$ (4.11) that depends on the volatility of trade value $\Omega^2{}_C(t)$ (4.12), volatility of trade volume $\Omega^2{}_U(t)$ (4.12), and their correlation $corr\{C(t)U(t)\}$ (4.13) (Olkhov, 2021; 2022):

$$\sigma^2(t) = \frac{\Omega_C^2(t) + a^2(t;1)\Omega_U^2(t) - 2a(t;1)corr\{C(t)U(t)\}}{U(t;2)} \geq 0 \quad (4.11)$$

Volatilities of trade value $\Omega^2{}_C(t)$ and volume $\Omega^2{}_U(t)$ (4.12) depend on the first two statistical moments of trade value $C(t;1)$, $C(t;2)$ and of trade volume $U(t;1)$, $U(t;2)$ (4.5):

$$\Omega_C^2(t) = C(t;2) - C^2(t;1) \quad ; \quad \Omega_U^2(t) = U(t;2) - U^2(t;1) \quad (4.12)$$

$$corr\{C(t)U(t)\} = E\big[\big(C(t_i) - C(t;1)\big)\big(U(t_i) - U(t;1)\big)\big] = E[C(t_i)U(t_i)] - C(t;1)U(t;1) \quad (4.13)$$

The relations (4.14) determine the joined average $E[C(t_i)U(t_i)]$ of trade value and volume:

$$E[C(t_i)U(t_i)] \sim \frac{1}{N}\sum_{i=1}^{N} C(t_i)U(t_i) \quad (4.14)$$

The 2-d market-based price statistical moment $a(t;2)$ takes the form:

$$a(t;2) = \frac{C(t;2) + 2a^2(t;1)\Omega_U^2(t) - 2a(t;1)corr\{C(t)U(t)\}}{U(t;2)} \quad (4.15)$$

One can show that for the case $U(t_i)=const$, $i=1,\ldots N$, the market-based price volatility $\sigma^2(t)$ (4.11) as well as VWAP $p(t;1)$ (4.6) coincide with the expressions of the price volatility and average price estimated by the frequency-based probability (4.3). That highlights the difference between the conventional frequency-based and market-based assessments of average price, price volatility, and price probability. The market-based assessments of price statistical moments take into consideration the randomness of market trade volumes. A market-based approach estimates the probability of a particular trade being proportional to its relative trade volume. Contrary to it, conventional frequency-based assessments assume that all market trades have equal probability $\sim 1/N$, regardless of the size of their market trade volumes. Actually, time series of real market trades reveal significant variations in trade volumes. Those who want to use the price statistical moments of the real market should follow the market-based approach to price probability.

The main outcome of our consideration of the market-base price volatility $\sigma^2(t)$ (4.11) is the highlight of its hidden complexity as the result of the dependence on volatilities (4.12) and correlation (4.13) of market trade values and volumes. The relations (4.10-4.15) tie up the properties of a random market trade with price volatility (4.11). Simply put, any predictions of market price probabilities at horizon $T$, even the predictions of the simplest Gaussian distributions that are determined only by price average $a(t;1)$ (4.6) and volatility $\sigma^2(t)$ (4.11), require forecasts of market trade volatilities (4.12) and correlation (4.13) at the same horizon $T$. That strictly links the projections of the price probabilities with the forecasting of market trade volatilities (4.12) and correlation (4.13). Moreover, as we argue in the next section, these links reveal the dependence on macroeconomic evolution as well.

## 5. Second-order economic theory

As we discussed above, the sums of market trade values and volumes of the first degree $C_{\Sigma}(t;1)$ and $U_{\Sigma}(t;1)$ (4.5) define the evolution of most macroeconomic variables. Macroeconomic investment and credits are composed of the sums of agents' investments and credits during $\Delta$ (4.2). In turn, the change in investment and credit of each agent during $\Delta$ is determined by the sums of the corresponding deals of each agent in investment and credit during $\Delta$. The sums of market trade values $C_{\Sigma}(t;1)$ and volumes $U_{\Sigma}(t;1)$ (4.7) during $\Delta$ (4.2) determine the origin of the first-order economic theory. The complexity of the evolution of the sums of market trades $C_{\Sigma}(t;1)$ and $U_{\Sigma}(t;1)$ (4.7) reveals the hidden challenges for macroeconomic forecasting.

Economic agents make their trade decisions in accordance with their expectations. The projections of the average price and, what is more important, the predictions of the price volatility $\sigma^2(t)$ (4.11) at a time horizon $T$ play the core role in setting agents' trade expectations. Thus, the dependence of price volatility $\sigma^2(t)$ (4.11) on volatilities of market trade values $\Omega^2_C(t)$, volumes $\Omega^2_U(t)$ (4.12), and correlations $corr\{C(t)U(t)\}$ (4.13) reveals the hidden influence of the sums of squares of market trades $C_{\Sigma}(t;2)$ and $U_{\Sigma}(t;2)$ (4.7) on the evolution of the first-order macroeconomic variables.

That is a very important and complex challenge for theoretical economics: the market-based price volatility $\sigma^2(t)$ (4.11) causes that the forecasts of the first-order macroeconomic variables that are determined by the sums of market trades $C_{\Sigma}(t;1)$ and $U_{\Sigma}(t;1)$ (4.7) depend significantly on the predictions of the evolution of the sums of squares of market trades $C_{\Sigma}(t;2)$ and $U_{\Sigma}(t;2)$ (4.7). To forecast the evolution of the sums of squares of market trades $C_{\Sigma}(t;2)$ and $U_{\Sigma}(t;2)$ (4.7), one should model the second-order macroeconomic variables. Macroeconomic variables of the second order should be collected via agents' variables of the second order in a manner similar to the methodology presented by Fox et al. (2017). However, such a methodology doesn't exist now and should be developed.

We highlight the importance and complexity of considering a new set of agents' variables – economic and financial variables of the second order. The variables of the second order are composed of sums of squares of trade values and volumes performed by the agents. For example, the second-order investment should be composed of squares of investment deals performed by the agent during $\Delta$ (4.2). The second-order credits should be composed of squares of credits received by the agent during $\Delta$ (4.2). The sums of squares of market trade values and volumes establish new additional set of second-order economic and financial variables for each particular economic agent. Almost each agent's variable of the first order should have its image as a variable of the second order. That significantly complicates economic theory but responds to internal economic laws: The forecasts of market-based price volatility that depend on volatilities and correlations of trade values and volumes depend on agents' variables of the second order.

The mutual influence of macroeconomic variables of the first and second orders establishes a tough problem for theoretical economics. We call the approximation, which takes into consideration the evolution of macroeconomic variables of the first and second orders the second-order economic theory.

Actually, the first two market-based price statistical moments give only a Gaussian approximation of the price probability. To derive a more precise approximation, one could make an attempt to derive the third price statistical moment. The third market-based price statistical moment $a(t;3)$ describes the price skewness $Sk(t)$ (Olkhov, 2022):

$$Sk(t)\sigma^3(t) = a(t;3) - 3a(t;2)a(t;1) + 2a^3(t;1) \qquad (5.1)$$

However, to forecast the approximation of price probability that takes into account the price skewness (5.1), one should predict the evolution of the sums of the third degree of trade values $C_\Sigma(t;3)$ and volumes $U_\Sigma(t;3)$ (4.7). This will many times increase the complexity of economic theory and will require the development of third-order economic theory.

Any attempt to obtain the next level of accuracy for the description and forecasting of the price probability increases greatly the complexity of economic theory. This exponentially growing complexity safely hides any exact predictions of price probability at horizon $T$ from anyone. However, understanding the origin and nature of the problem can help develop successive approximations of price probability that would predict economic evolution.

## 6. Conclusion

Market description is a necessary part of any model of theoretical economics. The choice of agents' variables, market trades, and expectations as primary bricks for the successive approximations of economic processes gives a wide opportunity to use various theoretical models and math. The projections of market trades mostly rely on agents' expectations, which to a large extent depend on the forecasts of price average and volatility. Any exact predictions of price probability or probabilities of market trade values and volumes are impossible. In the best case, one could predict a few statistical moments of price that provide only an approximation of the price probability. The forecasts of the average price and price volatility give only a simple Gaussian approximation of the price probability. However, even the predictions of the average price and price volatility at time horizon $T$ are extremely complex. The market-based price volatility $\sigma^2(t)$ (4.11) depends on the volatilities and correlation of trade values and volumes. To predict them, one should develop the second-order economic theory that models the sums of squares of trade values and volumes and describes macroeconomic variables of the first and second orders. These tough problems for years will design the subject of theoretical economics and second-order economic theory in particular.